\documentclass[a4paper,12pt]{article}
\usepackage{epsfig}

\advance \topmargin by -\headheight
\advance \topmargin by -\headsep
\evensidemargin \oddsidemargin
\advance\hoffset by -3mm  

\begin{document}

{\par
\title{Solid-state NMR quantum computer with individual access to qubits 
and some its ensemble developments}\par
\author{K. A. Valiev and A. A. Kokin}\date{}\maketitle\par
Institute of Physics and Technology, Russian Academy of Sciences, 
Nakhimovskii pr. 36/1, Moscow 117218, Russia,                       
E-mail: qubit@mail.ivvs.ru\par
{\abstract{}\par
Here we made an analysis of the principles of a semiconductor NMR 
quantum computer and its developments. The known variant of an 
individual-access computer (B. Kane) and alternative solid-state 
bulk-ensemble approach versions allowing to avoid some 
difficulties in implementing the first variant are considered.\par
}\section*{Introduction}\par
Atomic nuclei with a spin quantum number $I = 1/2$ seem to be the 
natural candidates for qubits --- two-level quantum elements in 
quantum computers. The early NMR quantum computers approach 
suggested in 1997 by two teams of researchers independently \cite{1,2} 
and then confirmed in experiments \cite{3,4}. In this approach it was 
used several diamagnetic organic liquids whith individual 
molecules, having a small number of tied non-equivalent nuclear 
spins--qubits, acting as almost independent quantum computers. 
Such an effect is due to the fact that vigorous rotational and 
translational Brownian motion of molecules in a liquid to a great 
extent averages both intra- and intermolecular dipole--dipole 
nuclear spin interactions. Decoherence time of the spin states 
thus turns out to be relatively large (several seconds or more). 
Only scalar intramolecular interaction between nuclear spins 
remains not averaged. It is described by the Hamiltonian 
${\sum_{\mathrm{i}<\mathrm{j}}} I_{\mathrm{i}\mathrm{j}} \hat\mathbf{I}_{\mathrm{i}}\hat\mathbf{I}_{\mathrm{j}}$ where $I_{\mathrm{i}\mathrm{j}}$ is an interaction constant. In particular, 
for the chloroform molecule $^{13}\mathrm{C}^{1}\mathrm{HCl}_{3}$ this interaction is essential 
only for proton $^{1}\mathrm{H}$ and carbon isotope $^{13}\mathrm{C}$ and it is this 
interaction around which unitary quantum operations on qubits have 
been performed.\par
Since in liquids the nuclear spins are weakly coupled with the 
environment, we may restrict our consideration to nuclear spins of 
an individual molecule ({\it reduced }quantum ensemble) rather than deal 
with a huge number of nuclear spins of all molecules of the 
liquid. For finite temperatures, the reduced ensemble is in the 
{\it mixed }state described by a diagonal density matrix. The nonzero 
elements of this matrix are the {\it occupancies }of associated spin 
states. For such a system could perform quantum computation, it 
should be initialized prior to data entry. {\it Initialization }means 
the separation of lower-dimension blocks from the density matrix 
of the initial mixed state that must have properties similar to 
those of pure states (a pure-state diagonal matrix may have only 
one nonzero element). States described by such matrix are called 
effective or {\it pseudo-pure}. All unitary quantum computations are 
performed using these states. Several approaches to initialization 
have been reported \cite{1,2,4,5}.\par
In an ensemble computer, many molecules--minicomputers, being 
nearly independent one another, act in parallel and can thus be 
controlled by operations on the entire (macroscopic) {\it volume }of the 
liquid. These operations are well-known in high-resolution NMR 
techniques. Access to individual qubits is replaced by 
simultaneous access to related qubits in all molecules of a bulk 
ensemble. Computers of this type are called {\it bulk-ensemble }quantum 
computers. They, in principle, can operate at room temperature. 
However, for magnetic fields used in conventional NMR 
spectrometers (no more than 20 T) relative spin polarization, 
specifying signals from pseudopure states, is very small even at 
low temperatures. In addition, the NMR signal intensity 
exponentially drops with the number of qubits \cite{2}. Therefore, an 
organic-liquid quantum computer operating at room temperature 
cannot have much more than 10 qubits in a molecule. Estimates 
show, however, that a quantum computer of practical value, i.e. 
with the number of degrees of freedom (or the dimension of the 
Hilbert space of a set of qubits) larger than in conventional 
customary computers, must have more than 40 qubits (2$^{40} \sim  10^{12}$ 
conventional bits).\par
Another radically new and still unimplemented design was 
proposed in \cite{6}. It involves the formation of an artificial 
multiple-spin system with access to {\it individual }nuclear spins--qubits. It is suggested to use an semiconductor MOS structure on a 
$^{28}\mathrm{Si}$ spinless substrate into a thin layer of which $^{31}\mathrm{P}$ stable 
phosphorus isotopes, acting as donors, are implanted. These donors 
substitute for silicon atoms at the lattice sites, producing 
shallow impurity states with a large effective Bohr radius, and 
have a nuclear spin number $I = 1/2$. The number of donors or qubits 
in such artificial "molecules" can be very large.\par
In this work (see also our article \cite{7}) we analyze in details 
the principles underlying NMR quantum computer operation 
(discussed in \cite{6}) and give a some further development. We call 
attention to the advantages and disadvantages of the early 
version. Possible alternative ensemble designs of solid-state--based nuclear--spin quantum computers that exploit the same 
principles but are free of the above disadvantages are next 
discussed.\par
\section{Basic requirements for semiconductor structures}\par
In the discussed Kane's approach temperatures must be low enough 
for electrons of the donors occupy only the lower spin state in a 
magnetic field; i.e. $T \ll  2\mu _{\mathrm{B}}B/k$ where $\mu _{\mathrm{B}} = 9.27\cdot 10^{-24}\,\mathrm{J}/\mathrm{T}$ is the 
Bohr magneton, $B$ is the induction of an external magnetic field 
and $k = 1.38\cdot 10^{-23}\,\mathrm{J}/\mathrm{K}$ is the Boltzmann constant. For $B \leq  2\,\mathrm{T}$ we 
have $T \geq  0.1\,\mathrm{K}$ which is much lower than the temperature for 
freezing-out donor electron states. Hence, the donors will remain 
for long in the neutral ground orbital $S$ state ($D^{0}-$state).\par
Each donor atom having a nuclear spin must be fairly 
accurately located under "its" metal gate (gate $\mathbf{A})$ separated from 
the semiconductor by a thin insulator film (for example, 
several-nanometer-thick silicon oxide). Gates $\mathbf{A}$ form a linear 
array of arbitrary length and period $l$ (one artifical "molecule") 
(Fig. 1). With the aid of the electric field generated by gates $\mathbf{A}$ 
one can affects the electron density distribution near the nuclei 
in the ground state, thereby {\it individually }adjusting the resonance 
frequency of every nuclear spin defined by external magnetic field 
and hyperfine nuclear--electron interaction. Thus, quantum 
operations by selectively applying resonant radio-frequency (RF) 
pulses to the nuclear spins of particular donors become feasible.\par
In deciding on the location of the donor atoms under the gate, 
one should bear in mind that silicon surface is not perfectly 
smooth; it will always have irregularities like valleys and 
ridges. Accordingly, the near-surface electric field under the 
gates at depths on the order of several lattice constants (in 
silicon, the lattice constant is 0.54 nm) will also be randomly 
non-uniform. Therefore, the electric field around the donors can 
be effectively controlled if they are more then 10 nm away from 
the silicon surface.\par
The characteristic sizes of the semiconductor structure 
(Fig. 1) are in a {\it nanometer }range. Such structures are fabricated 
with the use of modern nanotechnology techniques, such as 
epitaxial growth, ultra-high-vacuum (UHV) scanning probe 
nanolithography based on tunnel or atomic-force microscopes 
(AFM) \cite{8} and electron-beam or X-ray lithography.\par
The interaction of the nuclear spins of the donors with the 
environment can be precluded if the silicon and silicon oxide are 
enough purified of $^{29}\mathrm{Si}$ isotope. This isotope has a spin $I = 1/2$ 
and is contained in amounts of 4.7\% in natural silicon. For the 
number of silicon atoms per $\mathrm{cm}^{3}$ about 5.0$\cdot 10^{22}$ and the 
characteristic sizes depicted in Fig. 1, one donor atom occupies a 
volume of 20$^{3}\,\mathrm{nm}^{3} = 8\cdot 10^{-18}\,\mathrm{cm}^{3}$. On the average, less than one 
$^{29}\mathrm{Si}$ isotope atom must be present in this volume. Hence, the 
isotopic purity of silicon should be as high as $\sim 2\cdot 10^{-4}\%$.\par
In general, III--V semiconductors are inapplicable to quantum 
computation at this approach, since they do not possess spinless 
isotopes. However, we assume that spinless semiconductor 
structures can be built around other Group IV elements. Natural 
germanium has only one spin isotope $^{73}\mathrm{Ge}$ ($I = 9/2)$ in amounts of 
7.76\%. Natural carbon has spin isotope $^{13}\mathrm{C}$ with a spin number 
$I = 1/2$ in quantities of 1.1\%. Therefore, spinless structures can 
be made of Ge, Si/Si$_{1-\mathrm{x}}\mathrm{Ge}_{\mathrm{x}}$, and SiC materials if they are 
appropriately purified.\par
The nuclear spins of $^{29}\mathrm{Si}$ and $^{13}\mathrm{C}$ isotope atoms, which produce 
shallow impurity states in the above structures, can also be 
viewed as candidates for spins--qubits.\par
\section{Electron--nuclear spin system of a donor in a magnetic field}\par
Nuclear and electronic spins of donor atoms in the ground 
orbital state $\Psi _{0}(\mathbf{r})$ interact by hyperfine and dipole--dipole 
magnetic interactions. After averaging over the orbital electronic 
state, the latter vanishes, and only {\it hyperfine }interaction 
remains, whose Hamiltonian has the form (Fermi formula)\begin{eqnarray}
\hat{H}_{\mathrm{I}\mathrm{S}} = A \hat\mathbf{I}\hat\mathbf{S} ,\label{1}
\end{eqnarray}where\begin{eqnarray}
A = \frac{8\pi }{3} |\Psi _{0}(0)|^{2} 2\mu _{\mathrm{B}}g_{\mathrm{N}}\mu _{\mathrm{N}}\cdot \frac{\mu _{0}}{4\pi }\label{45}
\end{eqnarray}is hyperfine interaction constant, $\mu _{0} = 4\pi \cdot 10^{-1}\,\mathrm{T}^{2}\mathrm{cm}^{3}/\mathrm{J}$, 
$\mu _{\mathrm{N}} = 5.05\cdot 10^{-27}\,\mathrm{J}/\mathrm{T}$ is the nuclear magneton, and $g_{\mathrm{N}} = 2.26$ for 
$^{31}\mathrm{P}{\footnote
{$^{)}$ Hereafter, we use conventional designations $g_{\mathrm{N}},A$, and $J$ (for $J$ 
see below) instead of 2$g_{\mathrm{N}}$, 4$A$ and 4$J$, used in \cite{6}. 
}}^{)}$. With the experimentally found constant of hyperfine 
interaction $A = 7.76\cdot 10^{-26}\,\mathrm{J}$ (or $A/(2\pi \hbar ) = 116\,\mathrm{MHz})$ \cite{9} we obtain 
for the probability of an electron being on the donor nucleus 
($\mathbf{r} = 0)|\Psi _{0}(0)|^{2} = 0.43\cdot 10^{24}\,\mathrm{cm}^{-3}$.\par
The electron--nuclear spin Hamiltonian for a donor atom has 
the form\begin{eqnarray}
\hat{H} = 2\mu _{\mathrm{B}} \mathbf{B}\hat{\mathbf{S}} - g_{\mathrm{N}}\mu _{\mathrm{N}} \mathbf{B}\hat{\mathbf{I}} + A \hat\mathbf{I}\hat\mathbf{S} ,\label{2}
\end{eqnarray}four energy levels of which are given by the well-known Breit--Rabi formula. For $I = 1/2$ (the $z-$axis is parallel to $\mathbf{B})$ this 
formula is written as\begin{eqnarray}
E(F,m_{\mathrm{F}}) = - \frac{A}{4} - g_{\mathrm{N}}\mu _{\mathrm{N}}Bm_{\mathrm{F}} \pm  \frac{A}{2} \sqrt{1 + 2m_{\mathrm{F}}X + X^{2}} ,\label{3}
\end{eqnarray}where $X = (2\mu _{\mathrm{B}} + g_{\mathrm{N}}\mu _{\mathrm{N}})B/A,F = I \pm  1/2 = 1, 0$, and 
$m_{\mathrm{F}} = M + m = \pm  1, 0$ if $F = 1$ or $m_{\mathrm{F}} = 0$ if $F = 0$ (Here $M = \pm 1/2$ 
and $m = \pm 1/2$ are $z-$projections of electronic and nuclear spins 
accordingly). The energy levels vs. $X$ are shown in Fig. 2.\par
For the energy of the ground spin state, $F = 0$ and $m_{\mathrm{F}} = 0$ 
($M = -1/2,m = 1/2)$; hence, we obtain\begin{eqnarray}
E(0,0) = - A/4 - (1/2) \sqrt{(2\mu _{\mathrm{B}} + g_{\mathrm{N}}\mu _{\mathrm{N}})^{2}B^{2} + A^{2}}.\label{4}
\end{eqnarray}\par
For the next, excited energy state, $F = 1,m_{\mathrm{F}} = -1$, with the 
changed nuclear spin state, $M = -1/2,m = -1/2$, we have\begin{eqnarray}
E(1,-1) = A/4 - (2\mu _{\mathrm{B}} - g_{\mathrm{N}}\mu _{\mathrm{N}}) B/2.\label{5}
\end{eqnarray}\par
Thus, the energy difference between the two lower states of 
the nuclear spin that interacts with an electron whose state 
remains unchanged is described in simple terms ($\mu _{\mathrm{B}} \gg  \mu _{\mathrm{N}}$ for 
$A^{2}/(2\mu _{\mathrm{B}}B)^{2} \ll  1)$:\begin{eqnarray}
2\pi \hbar \nu _{\mathrm{A}} = E(1,-1) - E(0,0)
= A/2 + g_{\mathrm{N}}\mu _{\mathrm{N}}B + \frac{A^{2}}{8\mu _{\mathrm{B}}B} \sim  A/2 + g_{\mathrm{N}}\mu _{\mathrm{N}}B.\label{6}
\end{eqnarray}\par
For $^{31}\mathrm{P}$ donor atoms, the first term in (\ref{6}) exceeds the second 
one in fields $B < 3.5\,\mathrm{T}$. In this case, the nuclear resonant 
frequency $\nu _{\mathrm{A}}$ depends largely on the stationary local magnetic 
field $\mathbf{B}_{\mathrm{l}\mathrm{o}\mathrm{c}} = A\mathbf{S}/(g_{\mathrm{N}}\mu _{\mathrm{N}})$, which affects the nucleus via electron 
spin polarization $\mathbf{S}$. The external RF field $b_{\perp }(t)$, normal to the 
constant field $\mathbf{B}$, in NMR acts on nuclear spin not directly but 
through the transverse component of electronic polarization 
$B_{\mathrm{l}\mathrm{o}\mathrm{c},\perp } \sim  AS_{\parallel }b_{\perp }(t)/(g_{\mathrm{N}}\mu _{\mathrm{N}}B)$. This {\it gain effect }was first indicated by 
Valiev in \cite{10}.\par
Now, putting $B = 2\,\mathrm{T}$, according to \cite{6}, we obtain for the 
nuclear resonance frequency $\nu _{\mathrm{A}} = 92.6\,\mathrm{MHz}$.\par
\section{Electronic structure of the ground state of a donor and 
hyperfine interaction constant}\par
Consider the electronic structure of the ground state of a donor 
atom. The conduction band in silicon has {\it six }isoenergetic valleys 
with minima displaced relative to the Brillouin zone center by 
vectors $\mathbf{k}_{\mathrm{j}}$ ($j = 1,2,\ldots6)$ ($|\mathbf{k}_{\mathrm{j}}| = k_{0})$ toward three orthogonal axes 
of the fourth order. These axes are the axes of symmetry of the 
crystalline structure of silicon.\par
Let the direction to neighboring donors ($x-$axis) coincide with 
one of the axes of symmetry of the fourth order, for example 
[100], i.e., be perpendicular to the directions with whose four of 
the six energy-valley ellipsoids are aligned. Then, the effective 
Bohr radius $a$ of electrons for these {\it four }symmetric valleys in the 
indicated direction will be substantially larger than in any other 
direction, since this radius is defined mainly by the transverse 
effective mass $m_{t}$, which is much smaller than the longitudinal 
effective mass $m_{l}$. The value of the effective Bohr radius 
specifies the extent of the wave function of a donor electron and 
hence the characteristic scale of the semiconducting structure 
along $x-$axis.\par
Note that for the same reasons, in germanium, unlike silicon, 
the direction to a neighbor should be taken along one of six axes 
of symmetry of the second order, for example, [110]. This axis is 
perpendicular to four axes of the third order (in the Brillouin 
zone) along which the valley ellipsoids are aligned.\par
The orbital wave functions of an electron of a donor atom at 
the silicon lattice site are written in terms of tetragonal 
symmetry group representation $T_{\mathrm{d}}$ and are expressed by 
superpositions like \cite{11}\begin{eqnarray}
\Psi (\mathbf{r}) = {\sum_{\mathrm{j}=1}^{6}} \alpha _{\mathrm{j}} F_{\mathrm{j}}(\mathbf{r})\cdot \psi (\mathbf{k}_{\mathrm{j}},\mathbf{r}) ,\label{8}
\end{eqnarray}where $\psi (\mathbf{k}_{\mathrm{j}},\mathbf{r})$ is a Bloch function quite rapidly varying within a 
distance comparable to the lattice constant and corresponding to 
the six $\mathbf{k}_{\mathrm{j}}$ minima in the Brillouin zone. It is normalized to the 
unit cell volume $\Omega $:\begin{eqnarray}
\frac{1}{\Omega } \int_{\Omega } |\psi (\mathbf{k}_{\mathrm{j}},\mathbf{r})|^{2}d\mathbf{r}\mathbf{ }= 1.\label{9}
\end{eqnarray}\par
For silicon, it was found \cite{11} that $|\psi (\mathbf{k}_{\mathrm{j}},0)|^{2} = 186 \pm  18$, so 
a conduction electron is strongly localized at the lattice site. 
The wave functions $F_{\mathrm{j}}(\mathbf{r})$ describe smooth donor-related modulation 
of the Bloch functions and satisfy the Schr\"{o}dinger equation with 
effective masses\begin{eqnarray}
\left(- \frac{\hbar ^{2}}{2m_{l}} \frac{\partial ^{2}}{\partial z_{\mathrm{j}}^{2}} - \frac{\hbar ^{2}}{2m_{t}} \left(\frac{\partial ^{2}}{\partial x^{2}} + \frac{\partial ^{2}}{\partial y^{2}}\right) - \frac{q^{2}}{4\pi \epsilon _{\mathrm{s}}\epsilon _{0}r} - E_{\mathrm{d}}\right) F_{\mathrm{j}}(\mathbf{r}) = 0 ,\label{10}
\end{eqnarray}where the $z_{\mathrm{j}}-$axis is parallel to $\mathbf{k}_{\mathrm{j}}$ and originates at the donor 
atom; $E_{\mathrm{d}}$ are the energy eigenvalues of the donor electron; and 
$\epsilon _{\mathrm{s}} = 11.9 (\mathrm{Si}),\epsilon _{0} = 8.85\cdot 10^{-14}\,\mathrm{F}/\mathrm{cm}$.\par
The ground state of a donor electron in silicon belongs to 
one-dimensional (non-degenerate) representation $A_{1}$ of the tetragonal symmetry group $T_{\mathrm{d}}$. The wave function of this state can be 
represented as the superposition (\ref{8}) with equal weights $\alpha _{\mathrm{j}} = \sqrt{1/6}$ 
\cite{11}:\begin{eqnarray}
\Psi _{0}(\mathbf{r}) = \sqrt{1/6} {\sum_{\mathrm{j}=1}^{6}} F^{(1\mathrm{s})}_{\mathrm{j}}(\mathbf{r})\cdot \psi (\mathbf{k}_{\mathrm{j}},\mathbf{r}).\label{11}
\end{eqnarray}\par
For silicon and germanium, it was performed variational 
computations with the test modulating wave function \cite{11}\begin{eqnarray}
F^{(1\mathrm{s})}_{\mathrm{j}}(\mathbf{r}) = (\pi a^{2}_{t}b_{l})^{-1/2} \exp [ -(\rho ^{2}/a^{2}_{t} + z_{\mathrm{j}}^{2}/a^{2}_{l})^{1/2}] ,\label{12}
\end{eqnarray}where $\rho ^{2} = x^{2} + y^{2},a_{t}$ and $b_{t}$ are variational parameters for the 
ground state similar to the 1$s$ state of a hydrogen atom (the 
ground state under consideration turns into the 1$s$ hydrogen state 
if $m_{t }=_{ }m_{l})$. The obtained results somewhat different from the 
effective Bohr radii with the transverse and longitudinal masses:\begin{eqnarray}
a_{t}(\mathrm{Si}) = 2.50\,\mathrm{nm},&\,\,& a_{l}(\mathrm{Si}) = 1.42\,\mathrm{nm},\nonumber
\\
a_{t}(\mathrm{Ge}) = 6.45\,\mathrm{nm},&\,\,& a_{l}(\mathrm{Ge}) = 2.27\,\mathrm{nm}.\label{13}
\end{eqnarray}\par
The results for the ground state energy $E_{\mathrm{d}0}$ were the following 
(the values in parentheses stand for experimentally found 
ionization energies $E_{\mathrm{d}})$:\begin{eqnarray}
E_{\mathrm{d}0}(\mathrm{Si})&=&- 0.029\,\mathrm{eV} ( - 0.045\,\mathrm{eV}) ,\nonumber
\\
E_{\mathrm{d}0}(\mathrm{Ge})&=&- 0.009\,\mathrm{eV} ( - 0.012\,\mathrm{eV}).\label{14}
\end{eqnarray}\par
The states of the donor atom with the same function $F^{(1\mathrm{s})}_{\mathrm{j}}(\mathbf{r})$ 
but having different weight coefficients for two-dimensional, $E_{1}$, 
and three-dimensional, $T_{1}$, tetragonal group representations are 
excited and omitted from consideration.\par
The theoretically obtained probability of an electron being 
found in the vicinity of the donor nucleus in silicon,\begin{eqnarray}
|\Psi _{0}(0)|^{2} = 6 (F^{(1\mathrm{s})}_{\mathrm{j}}(0))^{2} |\psi (\mathbf{k}_{\mathrm{j}},0)|^{2} = \frac{6\cdot 186}{\pi a^{2}_{t}a_{l}} = 0,042\cdot 10^{24}\mathrm{cm}^{-3}\label{15}
\end{eqnarray}is almost 10 times less than the above defined experimental value, 
and the ionization energy is underestimated by a factor of 1.5. 
This is explained by the weaker coordinate dependence of the 
electrical potential in equation (10) produced by a donor atom in 
silicon at small distances intermediate between the covalent 
crystallochemical radius of phosphorus, 0.11 nm \cite{12}, and the 
silicon lattice constant, 0.54 nm. In this small region, one can 
put ($F^{(1\mathrm{s})}_{\mathrm{j}}(\mathbf{r}))^{2} \sim  (F^{(1\mathrm{s})}_{\mathrm{j}}(0))^{2} \neq (\pi a^{2}_{t}b_{l})^{-1}$. Using the experimental 
value for $|\Psi _{0}(0)|^{2}$, we find ($F^{(1\mathrm{s})}_{\mathrm{j}}(0))^{2} = 3.94\cdot 10^{20}\,\mathrm{cm}^{-3}$.\par
Consider now a variation of the ground state energy due to a 
deviation of the potential from the point-charge value \cite{13}:\begin{eqnarray}
\Delta E_{\mathrm{d}} = \frac{2\pi }{3} (F^{(1\mathrm{s})}_{\mathrm{j}}(0))^{2} \frac{q^{2}}{4\pi \epsilon _{\mathrm{s}}\epsilon _{0}} \overline{r^2} ,\label{16}
\end{eqnarray}where $\overline{r^2}$ is the average square of the radius of the region where 
potential deviation takes place. If the deviation of the ground 
state energy $\Delta E_{\mathrm{d}} = 0.045 - 0.029 = 0.016\,\mathrm{eV}$, we find from (\ref{16}) 
($\overline{r^2})^{1/2} = 0.42\,\mathrm{nm}$, quite a reasonable value.\par
It is noteworthy that, while computing $|\Psi _{0}(0)|^{2}$ for germanium 
with the use of correspondent variational parameters, one should 
take into account that the centers of isoenergetic ellipsoids are 
the centers of eight hexagonal Brillouin-zone faces; therefore, 
the sum over $j$ will contain only four terms (or four full 
ellipsoids), and the numerical value of $|\psi (\mathbf{k}_{\mathrm{j}},0)|^{2}$ will be 
different. For a $^{31}\mathrm{P}$ atom, the experimental value of the hyperfine 
interaction constant is $A = 15\cdot 10^{-4}\,\mathrm{cm}^{-1} = 45\,\mathrm{MHz}$, the singlet 
electronic state factor $g = 1.56$, and $|\Psi _{0}(0)|^{2} = 0,22\cdot 10^{24}\,\mathrm{cm}^{-3}$ 
\cite{9}.\par
\section{Hyperfine interaction constant of a donor atom versus electric 
field}\par
Consider in more details how the external electric field $E$ induced 
by the gate potential affects the hyperfine interaction constant 
A. We shall assume that the field is aligned with one of the axes 
of the fourth order. The origin, unlike equation (10), is now on 
the gate surface rather than at a donor atom (Fig. 1).\par
With regard to the fact that, at low temperatures, intrinsic 
semiconductor is essentially a dielectric and also considering 
that the thickness of substrate $D \gg  l_{\mathrm{A}} \sim  c \gg  d$ (Fig. 1), we shall 
express the electric field strength at a donor atom via the 
potential $V$ on gate $\mathbf{A}$, assuming that the gate is a circular disk 
of radius $a = l_{\mathrm{A}}/2$ lying on the surface of a semi-infinite 
dielectric. The expression for the electrical potential at a point 
with coordinates $\mathbf{r} = (\rho ,z),\rho ^{2} = x^{2} + y^{2}$ has the form \cite{14}\begin{eqnarray}
\varphi (\rho ,z) = \frac{2V}{\pi } \arctan  \sqrt{\frac{2a^{2}}{\rho ^{2}+ z^{2}- a^{2}+[(\rho ^{2}+ z^{2}- \mathrm{a}^{2})^{2}+ 4a^{2}z^{2}]^{1/2}}} ,\label{17}
\end{eqnarray}hence, for a line passing through the gate center $\rho  = 0$ near the 
donor atom, we have\begin{eqnarray}
\varphi (0,z) = \varphi (0,c) - E_{\mathrm{c}}(z - c) + E_{\mathrm{c}}^{\prime }(0,\mathrm{c})(z - c)^{2}/2 +\ldots ,\label{18}
\end{eqnarray}where\begin{eqnarray}
\varphi (0,c) = \frac{2V}{\pi } \arctan  \frac{a}{c} ,\nonumber
\end{eqnarray}\begin{eqnarray}
E_{\mathrm{c}} = E_{\mathrm{c}}(0,\mathrm{c}) &=& - \frac{d\varphi (0,c)}{dz} = \frac{2V}{\pi } \frac{a}{a^{2} + c^{2}} ,\label{19}
\\
E_{\mathrm{c}}^{\prime } = E_{\mathrm{c}}^{\prime }(0,\mathrm{c}) &=& \frac{dE(0,c)}{dz} = \frac{4V}{\pi } \frac{ac}{(a^{2} + c^{2})^{2}}\nonumber
\end{eqnarray}are the potential, the electric field strength and his gradient at 
the donor atom. Equation (10) should now be supplemented by the 
perturbed Hamiltonian $\hat{H}_{\varphi } = - q\varphi (0,z)$.\par
In the case of the almost homogeneous electric field in the 
neighborhood of the donor atom (contribution of term with $E_{\mathrm{c}}^{\prime }(0,\mathrm{c})$ 
is small if $ca^{*}_{\mathrm{B}}/a^{2} \ll  1,a^{*}_{\mathrm{B}} = 8\pi \epsilon _{0}\epsilon E_{\mathrm{d}0}/q^{2} -\,\mathrm{the}$ effective Bohr 
radius) a shift in the electronic density distribution relative to 
the position of the donor atom nucleus causes the same {\it decrease }in 
$A$, irrespective of the direction of the electric field with 
respect to $z-$axis. A correction to the wave function $F^{(1\mathrm{s})}_{\mathrm{j}}(c)$ 
modulating the ground state of a donor (its coordinates are now 
$\rho  = 0,z = c)$ is defined by the second-order expression in the 
perturbation theory. Non-diagonal matrix elements of the 
perturbation operator between even $F^{(1\mathrm{s})}_{\mathrm{j}}(\mathbf{r})$ states and odd excited 
states $F^{(2\mathrm{p})}_{\mathrm{j}}(\mathbf{r})$ and $F^{(3\mathrm{p})}_{\mathrm{j}}(\mathbf{r})$, similar to the 2$p$ and 3$p$ hydrogen 
states, are not zero. The energies of these excited donor states 
in silicon \cite{11} are $E_{2\mathrm{p}} = -0.0109\,\mathrm{eV}$ and $E_{3\mathrm{p}} = - 0.0057\,\mathrm{eV}$. Since 
the excited state functions $F^{(2\mathrm{p})}_{\mathrm{j}}(c) = F^{(3\mathrm{p})}_{\mathrm{j}}(c) = 0$, the desired 
correction, nonzero at $z = c$, is given only by that part of the 
second-order expression in the perturbation theory proportional to 
the ground state function $F^{(1\mathrm{s})}_{\mathrm{j}}(c)$ \cite{13}:\begin{eqnarray}
\Delta F^{(1\mathrm{s})}_{\mathrm{j}}(V) = - F^{(1\mathrm{s})}_{\mathrm{j}}(c) \frac{q^{2}E^{2}_{\mathrm{c}}}{2} {\sum_{\mathrm{m}}}^{\prime } \frac{|\langle 0|(z - c)|m\rangle |^{2}}{(E_{\mathrm{m}} - E_{\mathrm{d}})^{2}} ,\label{20}
\end{eqnarray}where $m = 2p$, 3$p$, etc. Since the energy differences between the 
ground and excited states, $E_{2\mathrm{p}} - E_{\mathrm{d}} = 0.034\,\mathrm{eV}$, 
$E_{3\mathrm{p}} - E_{\mathrm{d}} = 0.039\,\mathrm{eV}$, \ldots, 0.045 eV, are close to each other, we 
can estimate this correction through a second-order change in the 
ground state energy and then through the polarizability $\chi $  of the 
atom in an electric field by taking one of the factors in the 
denominator equal to some mean value $\delta E \sim  0.04\,\mathrm{eV}$ and carrying it 
out of the summation sign: 
\begin{eqnarray}
\Delta F^{(1\mathrm{s})}_{\mathrm{j}}(V) & \sim &  - F^{(1\mathrm{s})}_{\mathrm{j}}(c)\cdot \frac{q^{2}E^{2}_{\mathrm{c}}}{2\delta E} {\sum_{\mathrm{m}}}^{'} \frac{|\langle 0|(z - c)|m\rangle |^{2}}{E_{\mathrm{m}} - E_{\mathrm{d}}} =\nonumber
\\
& = & - 1/(2\delta E)\cdot (\chi E^{2}_{\mathrm{c}}/2)\cdot F^{(1\mathrm{s})}_{\mathrm{j}}(c).\label{21}
\end{eqnarray}\par
The polarizability $\chi $  of a donor atom in silicon can be 
estimated by the formula for a hydrogen atom \cite{13} using the 
effective Bohr radius $a^{*}_{\mathrm{B}} = 2\,\mathrm{nm}$, that leads to 
$\chi  = 4\pi \epsilon _{0}\cdot (9/2)(a^{*}_{\mathrm{B}})^{3} = 4\cdot 10^{-32}\,\mathrm{F}\cdot \mathrm{cm}^{2}$. Eventually, we have for 
correction (20)\begin{eqnarray}
\Delta F^{(1\mathrm{s})}_{\mathrm{j}}(V) = - 1.55\cdot 10^{-12}\cdot E^{2}_{\mathrm{c}}\cdot F^{(1\mathrm{s})}_{\mathrm{j}}(c),\label{22}
\end{eqnarray}where electric field $E_{\mathrm{c}}$ is given in V/cm. From (\ref{22}), it follows 
that the relative correction is small when $E_{\mathrm{c}} \ll  8\cdot 10^{5}\,\mathrm{V}/\mathrm{cm}$. For a 
field-dependent correction to hyperfine interaction constant, one 
obtains to second-order ($\sim  E^{2}_{\mathrm{c}})$ accuracy\begin{eqnarray}
\Delta A(V)/A \sim  - 3.1\cdot 10^{-12}\cdot E^{2}_{\mathrm{c}}.\label{23}
\end{eqnarray}\par
Substituting $E_{\mathrm{c}} = 0.25\cdot 10^{6} V$, found from (\ref{19}) for 
$c \sim  2a \sim  10\,\mathrm{nm}$, into (\ref{23}) yields $\Delta A(V)/A = - 0.19 V^{2}$ (here $V$ in V) 
or, for a $V-$dependent correction to resonance frequency,\begin{eqnarray}
\Delta \nu _{\mathrm{A}}(V) = \Delta A(V)/A\cdot \nu _{\mathrm{A}} = - 17.5\cdot V^{2}\,\mathrm{MHz}.\label{24}
\end{eqnarray}\par
This disagrees with results in \cite{6} where a frequency tuning 
parameter $\alpha  = d\Delta \nu _{\mathrm{A}}(V)/dV \Rightarrow  -30\,\mathrm{MHz}/\mathrm{V}$ at $V \Rightarrow  0$ is given. In 
general, the linear part of the $\Delta \nu _{\mathrm{A}}$ vs. $V$ dependence may be 
attributed to different causes. One of this is a non-homogeneity 
of the electric field (if $ca^{*}_{\mathrm{B}}/a^{2} \sim  1)$, and other is the built-in 
electric field in the semiconductor. In the first case it is 
necessary to take into account the last member of the expression 
(\ref{18}) which have the first-order perturbation theory nonzero value. 
The second case can take place when the work functions of the gate 
and substrate, which determine flat-band voltage $V_{\mathrm{F}\mathrm{B}}$, are 
distinguish. In this case, one should put $V^{2} \Rightarrow  (V_{\mathrm{F}\mathrm{B}} + V)^{2}$ in the 
above expressions. If for example, we put $V_{\mathrm{F}\mathrm{B}} \sim  0.6\,\mathrm{V}$,\begin{eqnarray}
\Delta \nu _{\mathrm{A}}(V) & \sim & - 17.5\cdot (V_{\mathrm{F}\mathrm{B}} + V)^{2} =\nonumber
\\
& = & - 6.3 - 21.0\cdot V - 17.5\cdot V^{2}\,\mathrm{MHz}.\label{25}
\end{eqnarray}\par
Note, that, as $V$ and $V_{\mathrm{F}\mathrm{B}}$ approach 1 V, the second-order 
expressions of the perturbation theory become poor approximation 
and corrections with $E^{3}_{\mathrm{c}}$ must be included in (\ref{22}).\par
The relative error in the hyperfine interaction constant $\overline{\delta A}/A$ 
due to technological inaccuracy $\overline{(\delta \rho)^2}$ in placing donor atoms in 
the $z-$axis is defined from (\ref{17}) and (\ref{23}) by the ratio $\sim $ 2$\overline{\delta \rho^2}/a^{2}$ 
for $c/a \ll  1$ and $\sim $ 2$\overline{\delta \rho^2}/c^{2}$ for $c \sim  a$. The condition $\overline{\delta A}/A \ll  1$ then 
specifies the {\it lower limit }for the parameters $c$ and $a \sim  l_{\mathrm{A}}/2$.\par
\par
\section{Electron--nuclear spin system for two interacting donor atoms}\par
The distance $l$ between donors, as well as the period of the 
semiconductor structure, is taken small enough. In this case, the 
constant $J$ of effective exchange electron--electron interaction 
$J(\hat\mathbf{S}_{\mathrm{a}}\hat\mathbf{S}_{\mathrm{b}})$ between two hydrogen-like neighbors a and b due to partial 
overlap of their electronic wave functions in the corresponding 
direction is the most sensitive to the field of the gate $\mathbf{J}$.\par
The total spin Hamiltonian for such a "molecule" is\begin{eqnarray}
\hat{H} = 2\mu _{\mathrm{B}}\mathbf{B}(\hat\mathbf{S}_{\mathrm{a}} + \hat\mathbf{S}_{\mathrm{b}}) + J \hat\mathbf{S}_{\mathrm{a}}\hat\mathbf{S}_{\mathrm{b}} + \Delta \hat{H} = \hat{H}_{0} + \Delta \hat{H},\label{26}
\end{eqnarray}where\begin{eqnarray}
\Delta \hat{H} = \hat{H}_{\mathrm{a}} + \hat{H}_{\mathrm{b}} = - g_{\mathrm{N}}\mu _{\mathrm{N}}B(\hat{I}_{\mathrm{z}\mathrm{a}}+ \hat{I}_{\mathrm{z}\mathrm{b}}) + A_{\mathrm{a}} \hat\mathbf{I}_{\mathrm{a}}\hat\mathbf{S}_{\mathrm{a}} + A_{\mathrm{b}} \hat\mathbf{I}_{\mathrm{b}}\hat\mathbf{S}_{\mathrm{b}}.\label{27}
\end{eqnarray}\par
In the absence of external magnetic field, for a positive 
value of the exchange interaction constant $J > 0{\footnote
{$^{)}$Often exchange interaction is written with the minus sign. We 
would then use designation $|J|$.}}^{)}$ ground state $^{1}\Sigma $  
 of the "molecule" is singlet. In \cite{6}, for well-separated hydrogen-like atoms, an asymptotic expression that does not include 
oscillations due to interference of the Bloch functions for 
various valleys was employed. In terms of the designations adopted 
in our work, it has the form \cite{22}{\footnote
{$^{)}$ This asymptotic expression differs from the known Sugiura's 
Heitler--London approximation, which gives a physically impossible 
positive value at extremely large $l/a_{t} > 49,5$.}}$^{)}$\begin{eqnarray}
J(l) \approx  1,6\cdot \frac{q^{2}}{4\pi \epsilon \epsilon _{0}a_{t}}\cdot (l/a_{t})^{5/2} \exp ( - 2l/a_{t}) ,\label{28}
\end{eqnarray}where $a_{t} = 3\,\mathrm{nm}$ for silicon (in \cite{6}, this variational parameter 
was used slightly lower: $a_{t} = 2.5\,\mathrm{nm})$.\par
According to variational parameters (\ref{13}), the extent of the 
wave function of a donor-atom electron, which is specified by $a_{t}$, 
in germanium turns out to be 2.6 times greater than in silicon. 
Accordingly, the distance at which the desired overlap of the wave 
functions of neighboring donors is achieved also increases. Thus, 
in the case of germanium, the scale of the semiconductor structure 
can be {\it considerably extended }if the axes of symmetry are properly 
oriented.\par
We shall omit for a while the interaction with nuclear spins 
in (\ref{26}), which here is a small perturbation, and after 
diagonalization of the Hamiltonian we arrive at three triplet 
($S = 1,M = 0,\pm 1)$ and one singlet ($S = 0,M = 0)$ energy levels for 
a dual-spin system ($M = M_{\mathrm{a}} + M_{\mathrm{b}} -\,\mathrm{the}$ projection of total 
electronic spin)\begin{eqnarray}
E_{0}(S,M) = J [S(S + 1)/2 - 3/4] + 2\mu _{\mathrm{B}}{\it B M }.\label{29}
\end{eqnarray}\par
Varying the potential of gates $\mathbf{J}$, located between gates $\mathbf{A}$, one 
can control the electronic density between neighboring donor atoms 
and thus the overlap of the wave functions of electrons localized 
on neighboring donors a and b, as well as the constant of their 
exchange interaction $J$ and the constant of indirect scalar 
interaction between their nuclear spins $I_{\mathrm{a}\mathrm{b}}$.\par
From Fig. 3, it follows that, for $J = 2\mu _{\mathrm{B}}B$, the singlet and 
one triplet levels are crossed as $B$ grows. In this range of the 
parameter $J$, even rather weak effects may essentially change the 
structure of the ground electronic state of the "molecule". The 
relationship $J(l) = 2\mu _{\mathrm{B}}B$ defines the necessary interdonor 
distance, which was estimated in \cite{6}. For $B = 2\,\mathrm{T}$, it is $l \sim  10-20\,\mathrm{nm} \gg  a_{t}$; hence, the {\it upper limit }for the size of gates $\mathbf{A}$ in such 
fields is $l_{\mathrm{A}} \sim  10\,\mathrm{nm}$. If germanium is used instead of silicon, 
rough estimates using the above values of the variational 
parameters yield much greater sizes: $l \sim  25-50\,\mathrm{nm}$ and $l_{\mathrm{A}} \sim  25\,\mathrm{nm}$.\par
Consider now hyperfine splitting of "molecular" levels due to 
the Hamiltonian of perturbation $\Delta \hat{H} (\ref{27})$ ($J \gg  A_{\mathrm{a}}, A_{\mathrm{b}})$. The main 
interest is the splitting of electronic levels, crossing and 
especially an anticrossing of states near the level crossing 
point.\par
We shall take into account, following on \cite{15}, that commutator 
$[ \hat{S}_{\mathrm{z}\mathrm{a}}+ \hat{S}_{\mathrm{z}\mathrm{b}} + \hat{I}_{\mathrm{z}\mathrm{a}}+ \hat{I}_{\mathrm{z}\mathrm{b}}, \hat{H} ] = 0$. Since the total electron-nuclear 
spin projection in the magnetic field direction 
$m_{\mathrm{a}}+ m_{\mathrm{b}}+ M_{\mathrm{a}}+ M_{\mathrm{b}} = m + M$ is conserved, matrix 16$\times 16$ that represents 
the total Hamiltonian, falls into five reduced matrices, 
corresponding to values $m + M = 0, \pm 1, \pm 2$. The four states, some 
of which are used for measurements of the nuclear spin states, 
correspond to $m + M = -1$.\par
We shall represent the perturbation Hamiltonian (\ref{27}) as the 
sum of the secular and nonsecular parts: $\Delta \hat{H} = \Delta \hat{H}_{\mathrm{s}\mathrm{e}\mathrm{c}} + \Delta \hat{H}_{\mathrm{n}\mathrm{s}\mathrm{e}\mathrm{c}}$, 
where\begin{eqnarray}
\Delta \hat{H}_{\mathrm{s}\mathrm{e}\mathrm{c}} = -g_{\mathrm{N}}\mu _{\mathrm{N}}B(\hat{I}_{\mathrm{z}\mathrm{a}}+ \hat{I}_{\mathrm{z}\mathrm{b}}) + (1/2)(A_{\mathrm{a}}\hat{I}_{\mathrm{a}\mathrm{z}}+ A_{\mathrm{b}}\hat{I}_{\mathrm{b}\mathrm{z}})(\hat{S}_{\mathrm{a}\mathrm{z}}+ \hat{S}_{\mathrm{b}\mathrm{z}}) ,\label{30}
\end{eqnarray}\begin{eqnarray}
[ \Delta \hat{H}_{\mathrm{s}\mathrm{e}\mathrm{c}}, \hat{H}_{0} ] = 0.\label{31}
\end{eqnarray}\par
The paired nuclear spins $I_{\mathrm{a},\mathrm{b}} = 1/2$ of the atoms a and b may 
be in three triplet ($I = I_{\mathrm{a}} + I_{\mathrm{b}} = 1,m = m_{\mathrm{a}} + m_{\mathrm{b}} = 0, \pm 1)$ and one 
singlet ($I = I_{\mathrm{a}} - I_{\mathrm{b}} = 0,m = 0)$ states.\par
States of the electronic--nuclear spin system will be 
designated as $|S,M; I,m\rangle $. They are eigenstates of the total 
Hamiltonian if the nonsecular part of hyperfine interaction $\Delta \hat{H}_{\mathrm{n}\mathrm{s}\mathrm{e}\mathrm{c}}$ 
is neglected.\par
Let us concentrate here on two cases:\par
1) For $J < 2\mu _{\mathrm{B}}B$, the lowest triplet state ($S = 1,M = -1)$ is 
the ground electronic state (Fig. 3). First-order energy 
corrections to (\ref{29}) within the perturbation theory depend on 
diagonal matrix elements of $\Delta \hat{H}$:\begin{eqnarray}
\Delta E_{1}(S,M; I,m) = \langle S,M; I,m|\Delta \hat{H}|S,M; I,m\rangle \label{32}
\end{eqnarray}\par
In this approximation, from the ground electronic level $E_{0}(1,-1)$ only three sublevels determined by the secular part of 
Hamiltonian are split:\begin{eqnarray}
\Delta E_{1}(1,-1; 1,-1) & = & g_{\mathrm{N}}\mu _{\mathrm{N}}B + (A_{\mathrm{a}}+A_{\mathrm{b}})/4 ,\nonumber
\\
\Delta E_{1}(1,-1;0,0) & = & \Delta E_{1}(1,-1;1,0) = 0 ,\label{33}
\\
\Delta E_{1}(1,-1; 1,1) & = & - g_{\mathrm{N}}\mu _{\mathrm{N}}B - (A_{\mathrm{a}}+A_{\mathrm{b}})/4 ,\nonumber
\end{eqnarray}\par
The states $|1,-1; 1,0\rangle $ and $|1,-1; 0,0\rangle $ that have the same 
value $m + M = -1$, remain degenerated. In the next approximation 
determined by the nonsecular part of hyperfine interaction, the 
lowest level $E_{0}(1,-1)$ in the electronic triplet for $J < 2\mu _{\mathrm{B}}B$ is 
split into four sublevels.\par
2) For $J > 2\mu _{\mathrm{B}}B$, the electronic singlet $|0,0\rangle $ is the ground 
electronic state. In the first approximation, the singlet ground 
electronic level is split into next three hyperfine sublevels{\footnote
{$^{)}$ In \cite{7} we have made mistakes for the values $\Delta E_{1}(0,0;0,0)$ and 
$\Delta E_{1}(0,0;1,0)$.}}$^{)}$:\begin{eqnarray}
\Delta E_{1}(0,0; 1,-1) & = & g_{\mathrm{N}}\mu _{\mathrm{N}}B ,\nonumber
\\
\Delta E_{1}(0,0;0,0) & = & \Delta E_{1}(0,0;1,0) = 0 ,\label{34}
\\
\Delta E_{1}(0,0; 1,1) & = & - g_{\mathrm{N}}\mu _{\mathrm{N}}B ,\nonumber
\end{eqnarray}\par
Among them one state $|0,0;1,-1\rangle $ have $M + m = -1$. If nonsecular
part of hyperfine interaction is accounted this state participates
in an anticrossing process for the state $|1,-1; 0,0\rangle $.\par
Near the crossing point $C$ for the ground electronic states the
nonsecular part of hyperfine interaction $\Delta \hat{H}_{\mathrm{n}\mathrm{s}\mathrm{e}\mathrm{c}}$ has next four 
nonzero non-diagonal matrix elements between states with the same 
values $m + M$:\begin{eqnarray}
\langle 0,0; 0,0|\Delta \hat{H}_{\mathrm{n}\mathrm{s}\mathrm{e}\mathrm{c}}|1,-1; 1,1\rangle & = & -(A_{\mathrm{a}}+ A_{\mathrm{b}})/4 ,\nonumber
\\
\langle 0,0; 1,-1|\Delta \hat{H}_{\mathrm{n}\mathrm{s}\mathrm{e}\mathrm{c}}|1,-1; 1,0\rangle & = & (A_{\mathrm{a}}- A_{\mathrm{b}})/4 ,\label{35}
\\
\langle 0,0; 1,-1|\Delta \hat{H}_{\mathrm{n}\mathrm{s}\mathrm{e}\mathrm{c}}|1,-1; 0,0\rangle & = & (A_{\mathrm{a}}+ A_{\mathrm{b}})/4 ,\nonumber
\\
\langle 0,0; 1,0|\Delta \hat{H}_{\mathrm{n}\mathrm{s}\mathrm{e}\mathrm{c}}|1,-1; 1,1\rangle  & = & (A_{\mathrm{a}}- A_{\mathrm{b}})/4.\nonumber
\end{eqnarray}\par
They describe the splitting of the hyperfine states that 
intersect in the absent of $\Delta \hat{H}_{\mathrm{n}\mathrm{s}\mathrm{e}\mathrm{c}}$. In order to find out the 
picture of the anticrossing process we shall choose then as basis 
the next four states of the electron--nuclear spin system with 
$M + m = -1:|S,M; I,m\rangle  = |1,-1; 0,0\rangle ,|1,-1; 1,0\rangle ,|1,0; 1,-1\rangle $ and 
$|0,0; 1,-1\rangle $. The reduced Hamiltonian can be represented then by 
matrix\begin{eqnarray}
\hat{H}_{0} = \left(
\begin{tabular}{c c c c }
$ J/4-2\mu _{\mathrm{B}}B$  & $(A_{\mathrm{a}}-A_{\mathrm{b}})/4$ & $-(A_{\mathrm{a}}-A_{\mathrm{b}})/4$ & $ (A_{\mathrm{a}}+A_{\mathrm{b}})/4$\\
$ (A_{\mathrm{a}}-A_{\mathrm{b}})/4$  & $J/4-2\mu _{\mathrm{B}}B$ & $ (A_{\mathrm{a}}+A_{\mathrm{b}})/4$  & $-(A_{\mathrm{a}}-A_{\mathrm{b}})/4$\\
$-(A_{\mathrm{a}}-A_{\mathrm{b}})/4$ & $(A_{\mathrm{a}}+A_{\mathrm{b}})/4$  & $ g_{\mathrm{N}}\mu _{\mathrm{N}}B+J/4$ & $-(A_{\mathrm{a}}-A_{\mathrm{b}})/4$\\
$ (A_{\mathrm{a}}+A_{\mathrm{b}})/4$ & $(A_{\mathrm{a}}-A_{\mathrm{b}})/4$ & $-(A_{\mathrm{a}}-A_{\mathrm{b}})/4$ & $g_{\mathrm{N}}\mu _{\mathrm{N}}B-3J/4$\\
\end{tabular}
\right)\label{36}
\end{eqnarray}\par
If for simplicity to assume here $A_{\mathrm{a}} = A_{\mathrm{b}} = A$, the matrix 
eigenvalues can be expressed by the simple equation:\begin{eqnarray}
\left[(J/4 - 2\mu _{\mathrm{B}}B - E)\cdot (g_{\mathrm{N}}\mu _{\mathrm{N}}B - J/4 -E) - A^{2}/4\right]^{2} -\nonumber
\\
- \left[(J/4 - 2\mu _{\mathrm{B}}B - E)\cdot J/2\right]^{2} = 0.\label{37}
\end{eqnarray}\par
In this case Hamiltonian is symmetric, as in the absence of 
hyperfine interaction, with respect to interchanging donors a and 
b and eigenstates of Hamiltonian are divided into two symmetric 
($I, S = 1)$ and two antisymmetric ($I + S = 1)$ states, that are 
correspondent to the superpositions of the paired states $|1,-1; 1,0\rangle ,|1,0; 1,-1\rangle $ and $|1,-1; 0,0\rangle ,|0,0; 1,-1\rangle $ with 
eigenvalues \cite{15}:\begin{eqnarray}
E_{\pm }^{\mathrm{s}} & = & g_{\mathrm{N}}\mu _{\mathrm{N}}B/2 + J/4 - \mu _{\mathrm{B}}B \pm  \sqrt{(\mu _{\mathrm{B}}B + g_{\mathrm{N}}\mu _{\mathrm{N}}B/2)^{2}+ (A/2)^{2}} ,\nonumber
\\
E_{\pm }^{\mathrm{a}} & = & g_{\mathrm{N}}\mu _{\mathrm{N}}B/2 - J/4 - \mu _{\mathrm{B}}B \pm  \sqrt{(\mu _{\mathrm{B}}B + g_{\mathrm{N}}\mu _{\mathrm{N}}B/2 - J/2)^{2}+ (A/2)^{2}}.\label{38}
\end{eqnarray}\par
The splitting of the states $|1,-1; 1,0\rangle $ and $|1,-1; 0,0\rangle $ 
degenerated in the first approximation is now\begin{eqnarray}
E_{-}^{\mathrm{s}} - E_{-}^{\mathrm{a}} & = & 2\pi \hbar \nu _{\mathrm{J}} \sim  \frac{(A/2)^{2}}{2\mu _{\mathrm{B}}B - J} - \frac{(A/2)^{2}}{2\mu _{\mathrm{B}}B} ,\,\,\,\, \mathrm{for}\,\,J \ll  2\mu _{\mathrm{B}}B ,\nonumber
\\
E_{-}^{\mathrm{s}} - E_{-}^{\mathrm{a}} & = & 2\pi \hbar \nu _{\mathrm{J}} \sim  A/2 ,\,\,\,\,\,\,\,\,\,\,\,\,\,\,\,\,\,\,\,\,\,\,\,\,\,\,\,\,\,\,\,\,\,\,\,\,\,\,\,\,\,\,\,\,\,\, \mathrm{for}\,\,J = 2\mu _{\mathrm{B}}B.\label{39}
\end{eqnarray}\par
The energy 2$\pi \hbar \nu _{\mathrm{J}}$ is the energy of transition of paired nuclear
spins from the upper triplet state to the singlet state. It is
known that such splitting can be described by spin Hamiltonian
like $I_{\mathrm{a}\mathrm{b}} \hat\mathbf{I}_{\mathrm{a}}\hat\mathbf{I}_{\mathrm{b}}$, where $I_{\mathrm{a}\mathrm{b}} = 2\pi \hbar \nu _{\mathrm{J}}$ is the indirect "exchange" 
integral for nuclear spins. For $B = 2\,\mathrm{T}$ and 
$J/2\pi \hbar  = 30\,\mathrm{GHz} < 2\mu _{\mathrm{B}}B/2\pi \hbar  = 57\,\mathrm{GHz},\nu _{\mathrm{J}}$ was estimated at 
$\nu _{\mathrm{J}} = 75\,\mathrm{kHz} \ll  \nu _{\mathrm{A}} = 92.6\,\mathrm{MHz}$ \cite{6}. Note that the frequency $\nu _{\mathrm{J}}$ grows 
as we approach the crossing point of the unperturbed levels $E_{0}(1,-1)$ and $E_{0}(0,0)$ and vanishes at $J \Rightarrow  0$.\par
As it follows from (\ref{38}), the electron-nuclear antisymmetric 
state $|1,-1; 0,0\rangle $ for $J < 2\mu _{\mathrm{B}}B$ undergoes the transformation into 
anticrossed antisymmetric state $|0,0; 1,-1\rangle $ for $J > 2\mu _{\mathrm{B}}B$. That is, 
the electron and nuclear subsystems exchange their states with the 
{\it opposite transfer of nuclear spin states }near the crossing point. 
The electron-nuclear symmetric state $|1,-1; 1,0\rangle $ for $J < 2\mu _{\mathrm{B}}B$ does 
not anticross with any ground electronic states $|0,0\rangle $ for $J > 2\mu _{\mathrm{B}}B$ 
and does not change nuclear spin states. Note that the splitting 
of symmetric and antisymmetric states levels for $J \gg  2\mu _{\mathrm{B}}B$\begin{eqnarray}
E_{-}^{\mathrm{s}} - E_{-}^{\mathrm{a}} = 2\pi \hbar \nu _{\mathrm{J}} \sim  J - 2\mu _{\mathrm{B}}B\label{40}
\end{eqnarray}is more then (\ref{39}) for $J \ll  2\mu _{\mathrm{B}}B$. Under these conditions, the 
constant of indirect scalar interaction $I_{\mathrm{a}\mathrm{b}}$ between the nuclear 
spins of the donor atoms a and b also radically changes.\par
The hyperfine splitting of triplet and singlet electronic 
levels near the point $C$ is shown schematically in Fig. 3.\par
\section{Measurement of nuclear spin states}\par
Measurement of individual nuclear spin states is one of the 
important problems. We will discuss here this very briefly.\par
The nuclear spin state is set for $J \ll  2\mu _{\mathrm{B}}B$, where nuclear 
spins are handled with NMR methods by applying RF pulses of 
resonance frequency. It is suggested to measure this state in two 
stages \cite{6}. Suppose that after quantum computation at $J \ll  2\mu _{\mathrm{B}}B$ two 
electrons, both on its donor atom, are initially in the triplet 
state that is in the $|1,-1\rangle $ state, the whole electron-nuclear 
system is either in $|1,-1; 1,0\rangle $ or in $|1,-1; 0,0\rangle $ state. By 
{\it adiabatically }increasing of the exchange parameter to $J > 2\mu _{\mathrm{B}}B$ in 
a process of the crossing point passage we lead to the electron-nuclear system transition from one antisymmetric state $|1,-1; 0,0\rangle $ 
to other $|0,0; 1,-1\rangle $ at the same total spin projection and allow 
{\it transfer the information }from nuclear to the electron spin 
subsystem.\par
In addition, if the energy of bound for an electron at one 
neutral donor (it is usually small) is more then its energy of 
attraction to the neighboring ionized donor ($D^{+}-$ state), the 
electron will be found near the neutral donor (D$^{-}-$state or helium-like atom) and charge transfer from one donor to the other will 
occur. This may be reach also by the corresponding change of the 
gate $\mathbf{A}$ electric potential. Therefore, {\it a charge transfer }from one 
donor to another takes place. It is supposed \cite{6} that this process 
can be detected with highly sensitive {\it single-electron capacitive 
techniques}.\par
It is important to note that the electrical measurements time 
should be significant less then the time of electron spin--lattice 
relaxation, that at low temperatures exceeds 10$^{3}\,\mathrm{s}$. In the case of 
an electrical control and electrical measurements the states of 
qubits the noise in electrical circuits are also significant. They 
cause fluctuations of the nuclear resonant frequencies $\nu _{\mathrm{A}}$ and $\nu _{\mathrm{J}}$ 
and lead in result to decoherence and dissipation of quantum 
states and to quantum errors in the process of calculations.\par
The detailed theoretical investigation of the electrical 
single-spin measurement method of electron and nuclear spin states 
by using single-electron transistors in simpler semiconductor 
structures with one double donor (in particular Te) was given 
in \cite{16}. Another scheme for measurement of the state of single 
spin based on the single-electron turnstile and injection of spin 
polarized electrons from magnetic metal contacts is proposed 
in \cite{24}.\par
\section{Bulk-ensemble solid-state quantum NMR computers}\par
The approach suggested in \cite{6} can solve a number of problems 
typical for the considered  individual-access computing. First, to 
initialize the starting state of the system, it will suffice to go 
to low temperatures $T \ll  2\mu _{\mathrm{B}}B/k$, at which each electronic spin of a 
donor atom is in the pure quantum ground state and thermal 
fluctuations are suppressed. Accordingly, the nuclear spins are 
also in the pure state. No special procedure to separate 
pseudopure states is required in this case. Second, the number of 
spins--qubits in this case is unlimited, which opens the door to 
creating an actual many-qubit quantum computer instead of 
demonstrating only quantum principles with the simplest spin 
systems on organic molecules. Third, there appears a possibility 
to detect, with single-electron techniques, electron transfer 
between adjacent $\mathbf{A}$ gates and thus to sense the state of individual 
nuclear spins. In \cite{6}, the possibility of creating hybrid 
quantum--classical systems was also noted. Here, quantum nuclear-spin devices are supplemented with conventional integrated 
circuits when many-qubit systems are difficult to obtain.\par
However, there are certain difficulties in implementing a 
quantum computer on individual donor atoms that was suggested in 
\cite{6}. These are, first of all, small signal from the spin of an 
individual atom, the need for use of high sensitive single-electron measurements, then the need for precision regular donor 
arrangement with nanometer scale that matches the gate chain, and 
at last the need for use of low temperatures.\par
As an alternative, we will briefly discuss here another 
feasibility of a bulk-ensemble silicon quantum computer. In this 
case, unlike the structure suggested in \cite{6}, gates $\mathbf{A}$ and $\mathbf{J}$ form a 
chain of narrow ($l_{\mathrm{A}} \sim  10\,\mathrm{nm})$ and long (several micrometers) strips 
along which donor atoms $L$ distant from each other are placed 
(Fig. 4). Thus, they form a quasi-one-dimensional regular 
structure. It can be considered also a random distribution of 
donors under gates.\par
We shall consider here only {\it two cases}.\par
Let $L$ be so much larger than $l$ that exchange spin interaction 
between electrons of donor atoms along the strip gates ($y-$axis) is 
negligibly small; i.e., $J(L), kT \ll  J(l), 2\mu _{\mathrm{B}}B$. Then, such a system 
breaks down into independent chains of the donor atoms in the 
direction transverse to the gates ($x-$axis). In that case, the 
regularity of donor structure along the strip gates does not play 
any role. The gates form an ensemble of independent equivalent 
multiple-qubit computers---artificial "molecules" whose electronic 
spins are initially aligned with the field. Accordingly, all 
nuclear spins--qubits are also similarly oriented. An output 
signal in this system, as in liquids, will be proportional to the 
number of the "molecules" or donor atoms under a strip gate $\mathbf{A}$. For 
example, for number of "molecules" more then thousand the length 
of the strips should be more then ten micron.\par
One would expect that with the bulk-ensemble approach, where 
many "molecules" work simultaneously, electrical measurements 
would be greatly simplified.\par
A more complicated situation appears when $L < l$. In this case, 
$J(L) \gg  2\mu _{\mathrm{B}}B, J(l) \gg  kT$, and exchange interaction between localized 
electronic spins along the strip gates will be favorable for 
producing one-dimensional antiferromagnetically ordered chains, 
that may be considered at low temperatures as a pure macroscopic 
quantum state. The neighboring electronic spins in the 
antiferromagnetic chain in the absence of the field are oppositely 
oriented. Due to hyperfine interaction in the ground state, 
nuclear spins will also be  oriented according to the electronic 
spin direction in the resultant field and will form in chains an 
antiferromagnetic type order.\par
The nuclear resonant frequencies $\nu _{\mathrm{A}}$ of neighboring nuclear 
spins will be different for each of the magnetic one-dimensional 
subarrays in the chain: 2$\pi \hbar \nu _{\mathrm{A}}^{\pm } \sim  |g_{\mathrm{N}}\mu _{\mathrm{N}}B \pm  \frac{A}{2}|$. As in the case of 
electronic spins oriented only in the external field, the RF field 
has a greater effect on a nuclear spin because of hyperfine 
interaction. A resulting NMR signal tuned only to one of the 
frequencies $\nu _{\mathrm{A}}^{\pm }$ will be proportional to half the number of nuclear 
spins in the antiferromagnetic chain or half the number of donor 
atoms under the gate $\mathbf{A}$.\par
The selective RF $\pi -$impulse tuned to one of the resonant 
frequency can invert the nuclear spins state of one subarray in a 
chain and the nuclear spins of whole chain will acquire quasi-one 
dimensional ferromagnetic ordering while the neighboring chain 
will remain in the antiferromagnetic state.\par
Let now the electronic spins of the such two neighboring 
chains are also setting for $J(l) \ll  2\mu _{\mathrm{B}}B$, that is let they and half 
the number of nuclear spins in the chains are in the state with 
the same orientations (triplet states of the spin pairs), while 
other half the number of nuclear spins is in the state with 
opposite orientations (singlet states of spin pairs). By 
adiabatically increasing of the exchange parameter by means of 
$\mathbf{J}-$gate potential varying to $J(l) > 2\mu _{\mathrm{B}}B$ in the passage of the 
crossing point we lead for each electron-nuclear pair of two 
neighboring chains to the transition from one antisymmetric state 
into other antisymmetric state. The triplet electronic pairs 
ground state of two neighbor chains must pass to the singlet 
state. The subarray of nuclear spins which is in the state with 
opposite orientation of nuclear spins relative to the subarray of 
neighboring chain will transit in the state with the same 
orientations. The electron subsystem of two chains simultaneously 
passes in antiphase state and at the same time to the half period 
shift of antiferromagnetically ordered electronic spins in the one 
chain relative to the other. The nuclear subsystem of the both 
chains becomes ferromagnetically ordered. Again, the Pauli 
principle allows two electrons to be now on the one site, if it is 
energetically profitable, that it allows transfer electrons 
between the neighboring chains and destruction the 
antiferromagnetic ordering in the chains. This process can be 
detected more easy than by means of highly sensitive single-electron spin states detecting methods.\par
The relaxation time $T_{2}$ of nuclear spins, which characterizes 
the {\it decoherence }time of their quantum states, depends largely on 
fluctuations of local fields that are determined mainly by 
interactions between nuclear and electronic spins of different 
atoms. If the times of electronic spin--spin and spin--lattice 
relaxations in a semiconductor ($\tau _{2}$ and $\tau _{1}$, respectively) are 
short, that play the correlation times of fluctuating fields, 
electron--nuclear dipole--dipole and scalar interactions will be 
strongly averaged, like in liquids, and the time $T_{2}$ will be large. 
Only the scalar part of indirect interaction between the nuclear 
spins of donors $I_{\mathrm{a}\mathrm{b}} \hat\mathbf{I}_{\mathrm{a}}\hat\mathbf{I}_{\mathrm{b}}$ will remain non-averaged.\par
However, when the states of individual nuclear spins are 
determined by electrical techniques, the time $\tau _{1}$ of electronic 
spin--lattice relaxation must be sufficiently large (several 
thousands of seconds) for the state of electronic spin have no 
time to relax during the electric measurements. This again 
indicates the need for low temperatures \cite{6,15}. In that case, for 
large enough distance between donors, large times $T_{1}$ and $T_{2}$ are 
provided mainly by the {\it small amplitudes }of fluctuating local 
hyperfine fields. In insulators at low temperatures this fields 
are determined by electron-phonon mechanism. In particular it was 
demonstrated \cite{17} that the phonon induced spin-lattice nuclear 
relaxation time $T_{1}$ for low temperatures in silicon is very large. 
In the case of antiferromagnetically ordering at low temperatures 
the fluctuating local fields are determined also by the 
interaction of nuclear spin with electron spin waves \cite{23}.\par
The decoherence is associated also with {\it noise voltages }across 
the gates $\mathbf{A}$. To suppress the decoherence of different origin ideas 
similar to those underlying high-resolution NMR methods may be 
used \cite{18}.\par
Upon forming structures with closely spaced and narrow gates 
$\mathbf{A}$, it would be more appropriate the lower arrangement of strip 
gates $\mathbf{J}$. In this case, gates $\mathbf{A}$ can be made wider for the same 
interdonor distance along the $x-$axis. If we completely do away 
with electrical measurements of individual states, the control 
gates $\mathbf{J}$ become unneeded and the semiconductor structure simpler. 
Scalar interaction between nuclear spins of donor atoms through 
their electronic states, as well as their resonant frequency, will 
be controlled only by gates $\mathbf{A}$ and external magnetic field.\par
An another interesting approach to implementation of a solid 
state {\it ensemble }quantum computer was described in \cite{19}, where it 
was considered a lattice of nuclear spins 1/2 having periodic 
structure of ABCABCABC\ldots type in two or three dimensions, where 
A,B,C--are nuclear spins only of {\it three types }with different 
resonance frequencies. It is supposed that the nuclei at three 
sites are embedded in a crystal lattice of some solid state 
compound with non-spinning nuclei and initialized with all spins 
at the ground $|0\rangle $ state. Each ABC--units of this superlattice can 
be used to store quantum information by setting one of spin up or 
down. This information can be moved around via some quantum 
cellular shifting mechanism. Cascading {\it unitary quantum }SWAP 
operations of A$\Leftrightarrow \mathrm{B}$, B$\Leftrightarrow \mathrm{C}$, C$\Leftrightarrow \mathrm{A}$, A$\Leftrightarrow \mathrm{B}$, \ldots, every which is achieved by 
cascading three quantum controlled NOT gates, can be used for this 
process. An ancillary donor nucleus D with spin 1/2 in the 
proximity of an A-site can serve as the input/output port.\par
{\it Universal }quantum logic is implemented with the aid of two-body interactions between two spins at D and a nearby site. A 
local environment region, about tens of ABC--units along all three 
dimensions provides a large quantum system with {\it a number of qubits 
over thousands }and only three types of nuclear spins. The whole 
crystal contains a {\it huge ensemble }of such identical NMR quantum 
computers---large artificial "molecules". The information in this 
case can be input by setting the D-spins to desired states then 
transferring to the nearest A-spins via the quantum SWAP 
operation. All these can be done selectively if the specified 
quantum transitions are driven by RF fields with distinguished 
frequencies. After this, the D-spins are reset to the $|0\rangle $ state.\par
The state of any qubits can be measured by moving it to the A-site nearest to the D-spins, then by swapping A$\Rightarrow \mathrm{D}$. Finally the 
state of the D-spins is measured by using NMR techniques. The 
proposed NMR quantum computer by the authors opinion can be easily 
scaled up and may work at low temperature to overcome the problem 
of exponential decay in signal-to-noise ratio. However, authors of 
the article \cite{19} did not made any numerical estimations and 
discuss any possibility of a concrete realization of this idea.\par
In conclusion, we will consider briefly possible nonelectric 
approaches that would allow ensemble solid state computers to 
actually operate at room temperature. Several such approaches are 
available from the literature.\par
Highly sensitive optical measurements of a change in 
Overhauser shift in exciton optical spectra were discussed in 
\cite{20}. The shift changed as a result of nuclear polarization in 
$\mathrm{GaAs}/\mathrm{Al}_{\mathrm{x}}\mathrm{Ga}_{1-\mathrm{x}}\mathrm{As}$ quantum dots. In-plane spatial resolution of 
$\sim  10\,\mathrm{nm}$ and a sensitivity of 10$^{4}$ nuclear spins at 6 K were 
achieved. This ensemble variant, based on optical detection, 
requires considerable redesign of the structure for laser 
radiation can strike required regions. Note that the sensitivity 
attained in \cite{20} is sufficient just for the ensemble approach. 
However, there are not well-known sensitive optical measurements 
for semiconductor structures in which silicon, germanium etc. are 
used.\par
At higher temperatures, a number of spins will pass into the 
excited state, while others will remain in the ground state. Of 
the latter spins, one can construct the pseudopure state, using 
methods mentioned in the Introduction for the ensemble approach. 
Another way to keep as many nuclear spins as possible in the 
ground state is to orient them not only with a permanent external 
magnetic field but also with dynamic methods, such as optical 
pumping \cite{21}, which make possible the orientation of nuclear spins 
as high as nearly 100\% even at room temperature. However, this can 
be done only for direct-gap semiconductors like III--V compounds. 
In the case of silicon or germanium this property of the band 
structure may be received, if the single crystals will be properly 
strained.\par
The authors are grateful to V.A.Kokin and L.E.Fedichkin for 
critical reading of the article and for useful remarks.\par
\par
 \pagebreak\section*{Figures}\par
\begin{center}\epsfbox{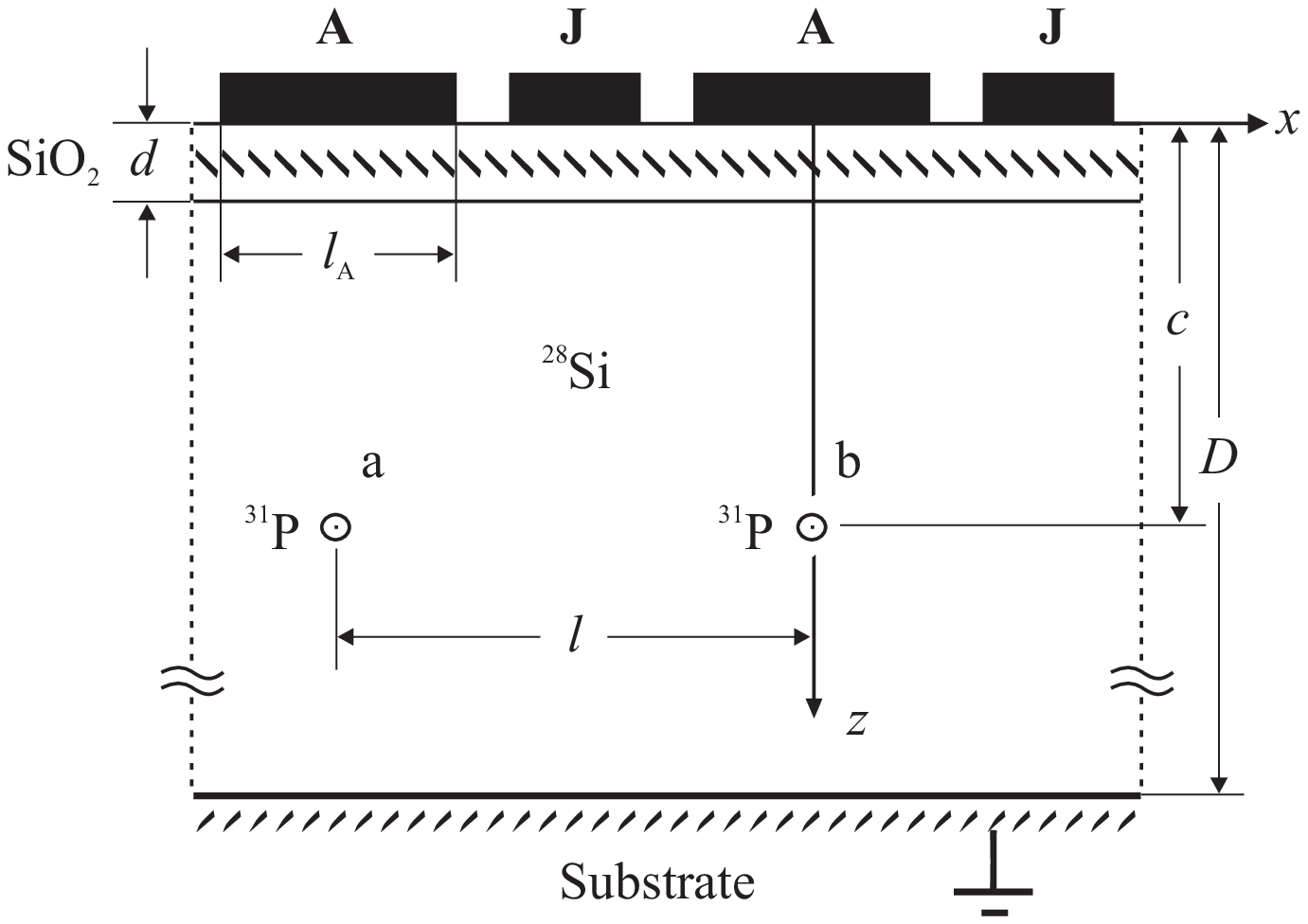}\end{center}\par Fig. 1. Two cells of the semiconducting structure with donor atoms 
a and b. $l_{\mathrm{A}} \sim  10\,\mathrm{nm},l \sim  20\,\mathrm{nm}$, and $c \sim  20\,\mathrm{nm}$ \cite{6}\par
\begin{center}\epsfbox{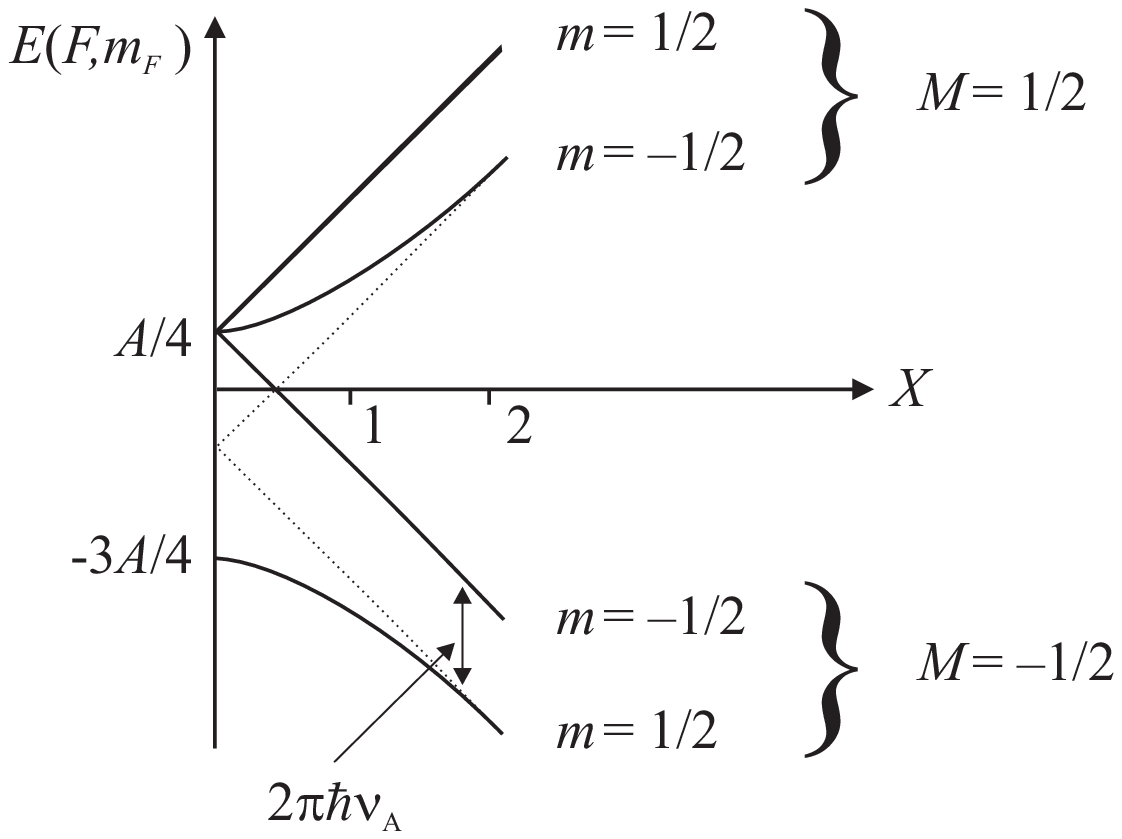}\end{center}\par Fig. 2. Energy levels in the electronic--nuclear system of an 
individual donor atom.\par
\begin{center}\epsfbox{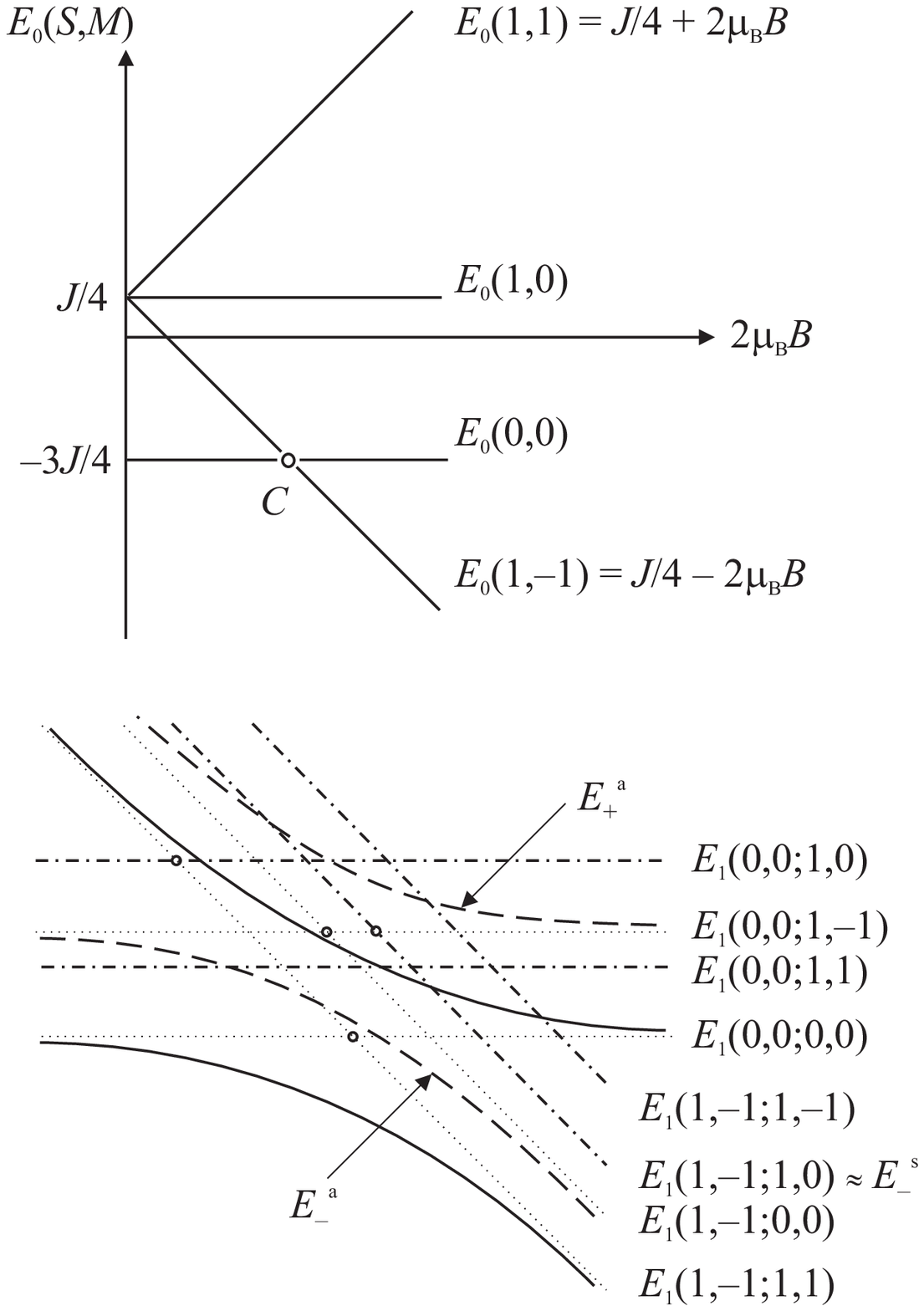}\end{center}\par Fig. 3. Energy levels of a dual-spin system subjected to a 
magnetic field in the absence of nuclear spins. The inset 
schematically shows hyperfine splitting of the ground 
electronic states near the crossing point $C$ for $A_{\mathrm{a}} \neq A_{\mathrm{b}}$. 
The circles indicate the anticrossing points of hyperfine 
states.\par
\begin{center}\epsfbox{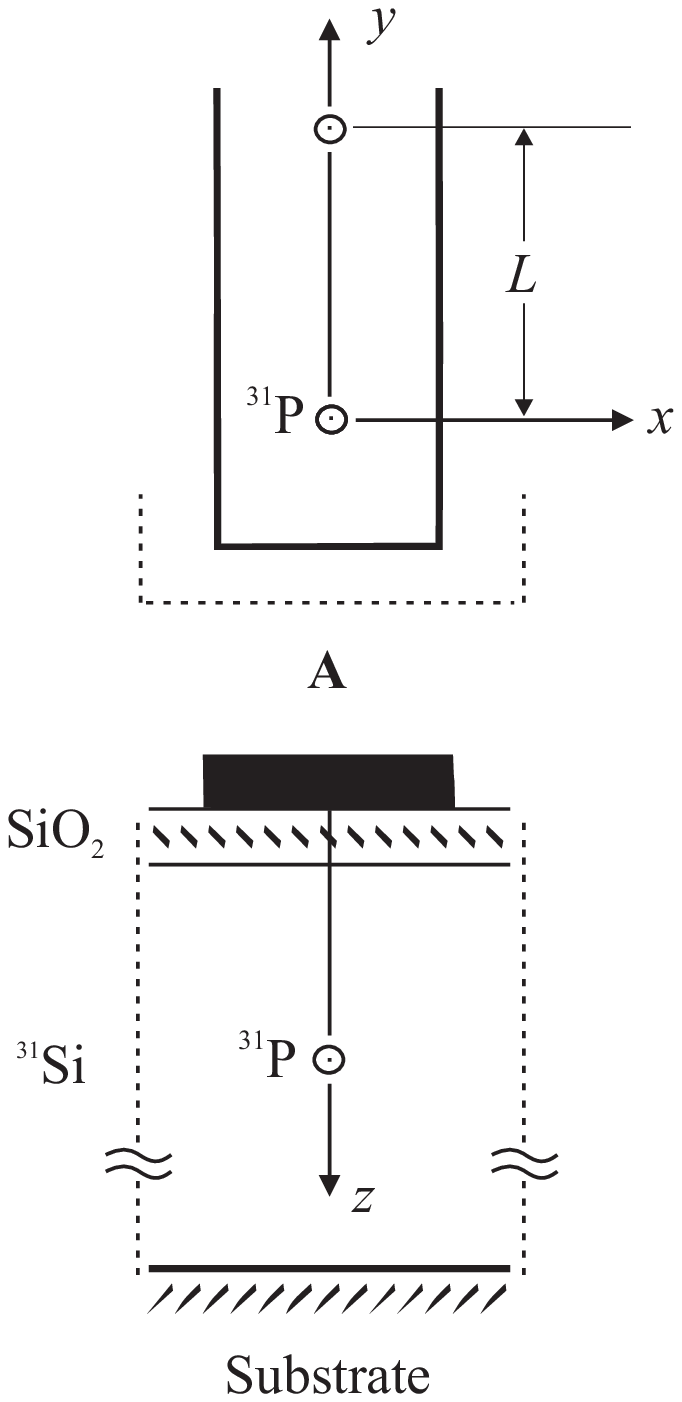}\end{center}\par Fig. 4. Fragment of the suggested ensemble structure (region only 
under gate $\mathbf{A})$.\par
}\end{document}